\documentclass[usegraphicx,usenatbib,useAMS]{mn2e}
\newcommand\lsim{\mathrel{\rlap{\lower4pt\hbox{\hskip1pt$\sim$}}
        \raise1pt\hbox{$<$}}}
\newcommand\gsim{\mathrel{\rlap{\lower4pt\hbox{\hskip1pt$\sim$}}
        \raise1pt\hbox{$>$}}}
\newcommand{\msun}{{\rm M_{\odot}}}

\usepackage{amsmath}
\usepackage{array}

\usepackage{graphicx}
\usepackage{fixltx2e}

\def\apj{ApJ}                 
\def\apjl{ApJL}               
\def\apjs{ApJS}               
\def\mnras{MNRAS}             
\def\aap{A\&A}                
\def\physrep{Phys.~Rep.}      
\def\nat{Nature}              

\title[Infrared Radiative Feedback]{Feedback from the IR Background in the Early Universe}
\author[J. Wolcott-Green and Z. Haiman]
{J. Wolcott-Green$^{1}$ and Z. Haiman$^{2}$ \thanks{E-mail: jw740@ast.cam.ac.uk; zoltan@astro.columbia.edu}\\ 
$^{1}$Institute of Astronomy, University of Cambridge, Madingley Road, Cambridge, CB3 0HA, U.K.\\
$^{2}$Department of Astronomy, Columbia University, 550 West 120th Street, MC 5246, New York, NY 10027, USA}
\begin{document}

\date{}

\pagerange{\pageref{firstpage}--\pageref{lastpage}} \pubyear{2012}

\maketitle

\label{firstpage}

\begin{abstract}
  It is commonly believed that the earliest stages of star--formation
  in the Universe were self-regulated by global radiation backgrounds
  -- either by the ultraviolet Lyman--Werner (LW) photons emitted by
  the first stars (directly photodissociating ${\rm H_2}$), or by the
  X-rays produced by accretion onto the black hole (BH) remnants of
  these stars (heating the gas but catalyzing ${\rm H_2}$ formation).
  Recent studies have suggested that a significant fraction of the
  first stars may have had low masses (a few ${\rm M_\odot}$). Such
  stars do not leave BH remnants and they have softer spectra, with
  copious infrared (IR) radiation at photon energies $\sim 1$eV.
  Similar to LW and X--ray photons, these photons have a mean--free
  path comparable to the Hubble distance, building up an early IR
  background.  Here we show that if soft--spectrum stars, with masses
  of a few ${\rm M_\odot}$, contributed $\gsim 1\%$ of the UV
  background (or their mass fraction exceeded $\sim 90\%$), then their
  IR radiation dominated radiative feedback in the early Universe.
  The feedback is different from the UV feedback from high-mass stars,
  and occurs through the photo-detachment of ${\rm H^-}$ ions,
  necessary for efficient ${\rm H_2}$ formation.  Nevertheless, we
  find that the baryon fraction which must be incorporated into
  low--mass stars in order to suppress ${\rm H_2}$--cooling is only a
  factor of few higher than for high-mass stars.
\end{abstract}

\begin{keywords}
cosmology: theory -- early Universe -- galaxies: formation --
molecular processes -- stars: Population III
\end{keywords}

\section{Introduction}

In hierarchical models of structure formation, the first stars in the
Universe form in dark matter (DM) minihalos with masses of $\sim
10^5~\msun$ at redshifts of $z\sim 20-30$, through efficient cooling
of the gas by ${\rm H_2}$ (\citealt{HTL96}; see a comprehensive review
by \citealt{BLreview}).  However, soon after the first stars appear,
early radiation backgrounds begin to build up, resulting in feedback
on star--formation.  In particular, the UV radiation in the
Lyman--Werner (LW) bands of ${\rm H_2}$ can photodissociate these
molecules and suppress gas cooling, possibly preventing star-formation
\citep{HRL97,ON99,HAR00,CFA00,MBA01,Ricotti+01,
  Ricotti+02b,MBH06,WA07,ON08,JGB08,WA08a,WA08b,Whalen+08,MBH09}.

Numerical simulations (e.g. \citealt{ABN02,BCL02,Yoshida+03}) have
long suggested that the metal--free stars forming in the early
minihalos were very massive ($\sim 100~\msun$), owing to the rapid
mass accretion enabled by ${\rm H_2}$ cooling. These stars would then
leave behind remnant BHs with similar masses \citep{Heger+03}, and
produce X-rays, either by direct accretion or by forming high-mass
X-ray binaries.  A soft X-ray background at photon energies of $\gsim
1$keV, at which the early intergalactic medium (IGM) is optically
thin, then provides further global feedback: both by heating the IGM,
and by catalyzing ${\rm H_2}$ formation in collapsing halos
\citep{HRL96,Oh01,Venkatesan+01,GB03,Madau+04,CM04,Ricotti+05,Mirabel+11}.

Recent simulations have been pushed to higher spatial resolution, and
in some cases, using sink particles, were able to continue their runs
beyond the point at which the first ultra-dense clump developed.  The
gas in the central regions of at least some of the early minihalos
were found to fragment into two or more distinct clumps
\citep{Turk+09,Stacy+10,Greif+11,Clark+11,Prieto+11}. This raises the
possibility that the first stars formed in multiple systems, and that
many of these stars had lower masses than previously thought (but see
\citealt{Turk+12} for still higher resolution simulations that suggest
less efficient fragmentation).

There is also some observational evidence suggesting a lower
characteristic Pop III mass.  Massive ($\gsim 140~{\rm M_\odot}$),
non-rotating metal-free stars are expected to end their lives as
pair-instability supernovae (PISNe), and the non--detection of the
characteristic PISN nucleosynthetic patterns in metal--poor stars
suggests that the typical Pop III stars did not form with such high
masses (see, e.g., a recent review by \citealt{FN11}). The
observations of carbon--enhanced metal poor stars may further imply a
significant number of Pop III stars with masses as low as $M=
1-8~\msun$ \citep{Tumlinson07a,Tumlinson07b}. Finally, the recent
discoveries of extremely metal poor stars with no sign of C or N
enhancement shows that low-mass star formation could occur at
metallicities much lower than previously assumed
\citep{Caffau+2011,Caffau+2012}, likely facilitated by dust
fragmentation \citep{Schneider+2012}.

Motivated by the above, in this {\em Letter}, we examine radiative
feedback from an early cosmic IR background, produced by a population
of low--mass stars.  Although we focus on low--mass PopIII (i.e.,
metal-free) stars, our conclusions are more generic, and apply at any
cosmic epoch when significant numbers of low--mass PopII stars
co-exist with massive PopIII stars.  Low-mass stars are expected to
have soft spectra, even in the metal-free case
\citep{TS00,Marigo+01,Schaerer02}, producing significant radiation at
$\sim 1$eV, near the photo-detachment threshold (0.76 eV) of the ${\rm
H^-}$ ion. ${\rm H^-}$ is a reactant in the dominant formation channel
for ${\rm H_2}$, (${\rm H^- + H \rightarrow H_2 + e^-}$) and its
destruction can therefore have a dramatic impact on the thermal
evolution of metal--free gas.

In fact, it is well known that photo-detachment of ${\rm H^-}$ by
cosmic microwave background (CMB) photons kept the ${\rm H_2}$
formation rate in the early Universe very low, until the CMB photons
redshifted to lower energies at $z \sim 100$ \citep{Hirasawa+69,HP06}.
${\rm H^-}$ photo-detachment can become globally important again once
stars begin to form, if they have soft spectra\footnote{To our
  knowledge, this point was first noted and discussed by
  \citet{Chuzhoy07}; see \S~\ref{subsec:others} below.}.

The aim of this {\em Letter} is to quantify (i) if and when, due to
low--mass stars, ${\rm H^-}$ photo-detachment again became the
dominant process to limit ${\rm H_2}$ cooling in the earliest
protogalaxies, and (ii) to what extent this may have increased or
decreased the net global negative radiative feedback in the early
Universe. We focus on the importance of this negative feedback for
minihalos (as opposed to the more massive halos that cool even in the
absence of molecular hydrogen). In order to accomplish this, we
perform ``one--zone'' calculations, following the coupled chemical and
thermal evolution of the gas in the presence of a cosmological
radiation background, including ${\rm H^-}$ photo-detachment by IR
photons and ${\rm H_2}$--photodissociation by LW photons.

The rest of this paper is organized as follows. In \S~\ref{sec:model}
we describe our chemical and thermal modeling. In \S~\ref{sec:results}, 
we present our results for the relative importance of IR radiative 
feedback, with various assumptions about the stellar populations; 
we also compare our results to previous studies.  Finally, in 
\S~\ref{sec:conclusions} we offer our conclusions. Throughout this 
paper we adopt a standard ${\rm \Lambda CDM}$ cosmological background 
model: ($\Omega_{\rm DM}, ~\Omega_{\rm b},~\Omega_{\Lambda},~h$)= 
(0.233, 0.046, 0.721, 0.701) \citep{Komatsu+11}.

\section{Modeling}
\label{sec:model} 
\subsection{Background Spectrum}

We assume that the early Universe is filled with background
radiation produced by stars.  Massive metal--free stars have hard
spectra, with effective blackbody temperatures of $T_{\rm eff}\approx
10^5$K on the zero-age main sequence (ZAMS), nearly independent of
their mass above $M_\star\gsim100~\msun$
\citep{Marigo+01,BKL01,Schaerer02}. Below this mass, the effective
temperature decreases monotonically, and reaches $T_{\rm eff}\approx
10^4$K at $M_\star\approx 2~\msun$ \citep{TS00,Marigo+01,Schaerer02}.

We first consider a composite spectrum produced by co-existing low--
and high--mass stars, the relative abundances of which are allowed to
vary. We choose characteristic masses of $M_{\rm lo}= 1.2~\msun$ and
$M_{\rm hi} = 100~\msun$, which have time--averaged effective
temperatures of $T_{\rm eff,lo}=10^{3.95}$K and $T_{\rm
eff,hi}=10^{4.88}$K, respectively \citep{Marigo+01}. These stars have
main sequence lifetimes of $t_{\rm ms}\approx 3$ Gyr and $\approx 3$
Myr.  In our analysis below, for low--mass stars with $t_{\rm
ms}<0.5$Gyr (the age of the universe at $z\approx 10$), we reduce the
total radiative output by the factor ($t_{\rm ms}/0.5$Gyr).  We
parameterize the population by the mass fraction of low--mass stars,
$f^{\rm mass}_{\rm lo}$ and also by their corresponding fractional contribution
$f^{\rm uv}_{\rm lo}$ to the total UV radiation output (at 13.6eV).

We caution that this is of course only a crude characterization of the
true background - in principle, one needs to consider the
time-evolving spectra of stars with a range of initial
masses. Departures from a black-body shape may also be important, as
our results are sensitive to photons with the particular energies of
$\sim 2$eV (for ${\rm H^-}$ detachment) and $\sim 12$eV (for ${\rm H_2}$ 
dissociation). However, given that the stellar IMF is unknown,
we here adopt this simple prescription and defer a more detailed
treatment to future work.  In order to understand the dependence on
stellar mass, in \S~\ref{sec:SingleT} we repeat our calculations
assuming the background is produced by stars with a single mass in the
range $0.7 \leq {\rm M_{char}}/\msun \leq 100$ (corresponding to
effective temperatures in the range $6000 \lsim T_{\rm eff} \lsim
10^5$K).

The early IGM is optically thin at the IR energies of $\sim 2$eV,
relevant for photo-detachment of ${\rm H^-}$ (owing to the very low
abundance of both intergalactic ${\rm H^-}$ and of $e^-$).  However,
before reionization, the IGM is opaque above 13.6eV, and we
accordingly assume zero flux above this energy.  UV photons in the LW
bands (11.2-13.6eV), traveling over cosmological distances, will also
be absorbed by HI once they redshift into resonance with a Lyman
line. This results in a ``sawtooth modulation'' \citep{HRL97}, which
reduces the ${\rm H_2}$ dissociation rate by about an order of
magnitude at $z=15$ compared to the optically thin rate \citep{WGH11}.
We include this reduction in our calculation.  In principle, the
optical depth in the LW lines due to intergalactic ${\rm H_2}$ itself,
with an abundance of $n_{\rm H2}/n_{\rm H}\sim 10^{-6}$, can be of
order a few \citep{HAR00,Ricotti+01,KM05}; this additional opacity
could further reduce the ${\rm H_2}$--dissociation rate and would
strengthen our conclusions; we conservatively ignore it in our
calculations.

\subsection{Chemistry and Cooling}

We model a static gas cloud which has condensed to the maximum density
achievable by adiabatically collapsing in a minihalo with virial
temperature $T_{\rm vir}$: $\rho_{\rm max}= f_\rho \rho_{\rm IGM}
(T_{\rm vir}/T_{\rm IGM})^{3/2}$. Here $\rho_{\rm IGM}$ and $T_{\rm
  IGM}$ are the density and temperature of the smooth background IGM,
and $f_\rho$ (chosen below) is of order unity and is used for
calibration against simulations. At this stage, only if sufficient
${\rm H_2}$ forms will the cloud be able to radiatively cool and
continue collapsing to higher densities, and ultimately form stars.

We follow the chemical and thermal evolution for primordial gas, using
a standard chemical reaction network among nine species: ${\rm H,~
H^+,~H^-,~ He,~ He^+, ~He^{2+},~ H_2,~ H_2^+}$ and $e^-$ (and
photons). Radiative cooling by ${\rm H_2}$ is modeled as in
\citet{GP98}.  Our chemical model is adopted from \citet{SBH10}, where
the reader is referred for more details and references.  We adopt
initial conditions appropriate for a minihalo (i.e. with molecular
hydrogen and electrons fractions of $x_{\rm H_2} = 2 \times 10^{-6}$
and $x_{\rm e} = 10^{-4}$, and initial gas temperature $T_{\rm
gas}=T_{\rm vir}=400$K).  We follow the coupled thermal and chemical
evolution using the stiff equation solver {\sc lsodar} for $\approx
10^8$ years (or $\sim 20$ per cent of the Hubble time at $z\sim 10$,
after which it is likely the halo will have merged with another to
form a larger collapsed object).

Photo-detachment of ${\rm H^-}$ by continuum photons (${\rm H^- +
\gamma \rightarrow H + e^-}$) is dominated by $\sim 2$eV photons (at
which the ${\rm H^-}$ detachment rate peaks for $T_{\rm eff}\approx
10^{4}$K, slightly above the 0.755eV threshold). We fit the
frequency--dependent cross-section as in \citet{SK87}, convolving this
with the blackbody spectrum to find the rate coefficient,
parameterized as $k_{25}=10^{-10} \alpha J_{21}~{\rm s^{-1}}$. Here
and below, $J_{21}$ denotes the specific intensity at the Lyman limit
(13.6eV) in units of $10^{-21}{\rm ~erg~s^{-1}
~cm^{-2}~sr^{-1}~Hz^{-1}}$. For the composite spectrum with $(T_{\rm
eff,lo}, T_{\rm eff,hi})=(10^{3.95}{\rm K},10^{4.88}{\rm K})$, we have
$\alpha=\alpha_{\rm lo}f^{\rm uv}_{\rm lo}+\alpha_{\rm hi} (1-f^{\rm
uv}_{\rm lo})$, and ($\alpha_{\rm lo},\alpha_{\rm hi}) = (10^4,0.17)$.
 
For the photodissociation of ${\rm H_2}$ by LW photons we use the
fitting formulae for the optically--thick rate provided by
\citet{WHB11}, which includes self--shielding as well as shielding by
HI. The column densities are specified by assuming the size of the
collapsing region equals the Jeans length; however, we reduce $N_{\rm
H_2}$ by a factor of 10, which produces better agreement with the
shielding found in three--dimensional simulations \citep{WHB11}.

\section{Results and Discussion}
\label{sec:results}

\subsection{Composite Spectrum}
\label{sec:Composite}

{\it The Critical Flux}\\ To assess whether ${\rm H_2}$--cooling can
be suppressed by the IR background, we first run one-zone models at
various different radiation intensities, normalized at the
Lyman--limit $J_{21}$. We employ a Newton-Raphson scheme to find the
``critical flux,'' $J_{\rm crit}$, defined by requiring that the
cooling time always remains longer than the dynamical time (eq.~10 in
\citealt{HRL97}). Repeating this calculation for halos with different
masses and redshifts, we have found that with the density
normalization $f_\rho=0.5$, our resulting $J_{\rm crit}=J_{\rm
  crit}(M_{\rm halo},z)$ agrees well (to within a factor of two) with
the values, and the mass-- and redshift--dependence of $J_{\rm crit}$
found in simulations \citep{MBA01,MBH06}.  Throughout this section, we
show our results for a single halo mass scale, $T_{\rm vir} = 400$K
and redshift $z=18$ (corresponding to the lowest-mass minihalos that
can cool via ${\rm H_2}$).  Apart from a monotonic overall increase of
$J_{\rm crit}$ with $T_{\rm vir}$ and with $z$, our results are
qualitatively the same for other redshift and halo masses (up to the
scale corresponding to $T_{\rm vir} \sim 10^4$K, at which atomic
cooling becomes important).

\begin{figure}
  \includegraphics[width=70mm]{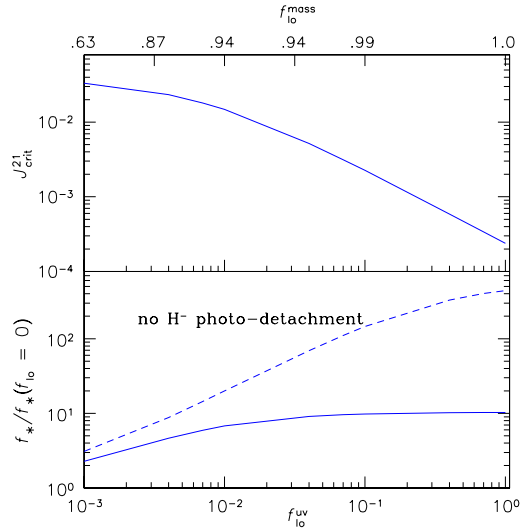}
  \caption{{\it Top panel:} The critical flux $J^{21}_{\rm crit}$
    required to suppress ${\rm H_2}$--cooling as a function of $f^{\rm
    uv}_{\rm lo}$.  Here $f^{\rm uv}_{\rm lo}$ is a proxy for the
    fraction of the Lyman limit flux contributed by low--mass stars
    with an effective temperature $T_{\rm eff} \approx10^4$K (rather
    than high-mass stars with $T_{\rm eff} \approx10^5$K). The
    corresponding mass fraction in the low--mass stars, $f^{\rm
    mass}_{\rm lo}$, is shown along the upper horizontal axis.  {\it
    Bottom panel:} The fraction of baryons that must be converted into
    stars in order to produce the critical flux shown in the top panel
    (solid line). Also shown is the required fraction if ${\rm H}^-$
    photo-detachment is artificially switched off in the chemistry
    network (dashed line).}
  \label{fig:CompositeSpec}
\end{figure}

Our main result is shown in Figure~\ref{fig:CompositeSpec}, the top
panel of which shows the critical flux $J_{\rm crit}$ as a function of
$f^{\rm uv}_{\rm lo}$ (the corresponding mass fraction $f^{\rm
  mass}_{\rm lo}$ is shown along the upper horizontal axis). The
critical flux decreases dramatically as $f^{\rm uv}_{\rm lo}$
increases above the per cent level. Since the normalization of the
critical flux is quoted at 13.6eV, close to the LW band, the critical
LW dissociation rate would be nearly independent of $f^{\rm uv}_{\rm
  lo}$; the decrease is caused entirely by the ${\rm H^-}$ detachment
by IR photons. The IR radiation becomes more important than the LW
radiation when the decrease reaches a factor of two, which occurs for
$f^{\rm uv}_{\rm lo}\approx 10^{-2}$.  However, because the low--mass
stars emit far fewer Lyman--limit photons than those with effective
temperatures $\sim 10^5$K, a mass fraction of $f^{\rm mass}_{\rm
  lo}\approx 0.9$ is required to achieve even this percent level
contribution to the flux (as shown by the upper horizontal axis).  For
reference, the mass fraction of stars in the 0.1-1.2${\rm M_\odot}$
(0.1-2${\rm M_\odot}$) range for the commonly used Chabrier IMF is
86\% (94\%).  In the limit of purely low-mass stars, $J_{\rm crit}$ is
decreased by a factor of $\approx 150$. (Note that \citealt{Clark+11}
do find the protostellar IMF peaked at $\lsim 1 \msun$, but ZAMS
masses are unknown.)

\noindent{\it Global Impact}\\ We next ask whether the critical flux
$J_{\rm crit}$=few$\times(10^{-4}-\times10^{-2})$ can be produced by
the mixture of high-- and low--mass stars. To answer this, we
determine the fraction of baryons that must be incorporated into stars
to produce a given $J_{\rm crit}$, as a function of $f^{\rm uv}_{\rm
lo}$. The energy density in LW photons $u_\gamma \simeq 4\pi\nu J_{\rm
crit}/c$ is converted to a stellar mass density $\rho_\star$ by
computing the number of LW photons emitted per stellar
baryon.\footnote{We assume a flat spectrum across the narrow LW bands;
the ``sawtooth modulation'' by the IGM has been absorbed into $J_{\rm
crit}$.}  We use the data from \citealt{Marigo+01} for this purpose,
and find the mass--weighted average for the high-- and low--mass
populations.

The bottom panel of Figure~\ref{fig:CompositeSpec} shows the critical
baryon fraction, $f_{\rm \ast,crit}$, which must be incorporated into
stars to achieve $J_{\rm crit} (f^{\rm uv}_{\rm lo})$ (normalized to
its value in the absence of any low-mass stars).  The fraction varies
relatively little over the range $0 \leq f^{\rm uv}_{\rm lo} \leq 1$
(solid curve).  The factor of $\sim 10$ increase as $f^{\rm uv}_{\rm
lo}\rightarrow 1$ is due almost entirely to the long lifetimes of the
low-mass stars, and the corresponding reduction of their radiation
output over the finite 0.5 Gyr age of the universe (at $z\approx10$).
The otherwise near-flatness of this curve is a coincidence: increasing
$f^{\rm uv}_{\rm lo}$ decreases both the critical flux and the
production of LW photons per stellar mass, and these two factors turn
out to nearly cancel. The bottom-line is that ${\rm H^-}$
photo-detachment can indeed suppress subsequent star formation; this
requires converting baryons into low-mass stars up to an order of
magnitude more efficiently than in the high-mass case.  Finally, the
dashed curve in the bottom panel in Figure~\ref{fig:CompositeSpec}
shows that if the IR photons from the low-mass stars were neglected,
the required stellar density would increase by two orders of
magnitude. This highlights the importance of including IR feedback in
future models if the Pop III IMF indeed extends to such low masses
(and also at all cosmic epochs when PopII stars are already present).

\subsection{Dependence on effective temperature}
\label{sec:SingleT}
\begin{figure}
  \includegraphics[width=70mm]{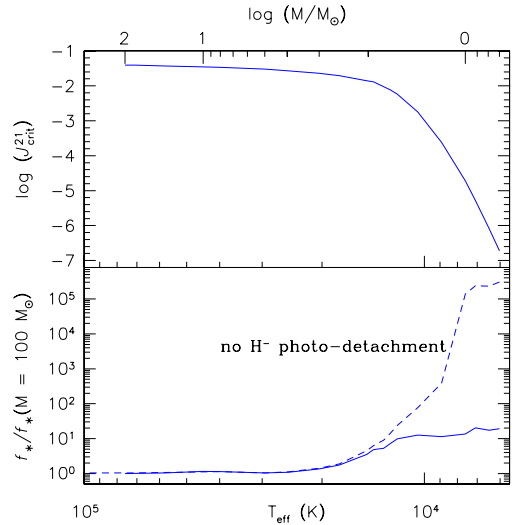}
  \caption{{\it Top panel:} Critical flux normalized (as in
    Fig.~\ref{fig:CompositeSpec}) at the Lyman limit, as a function of
    effective temperature of the incident blackbody radiation. The
    corresponding masses of zero--metallicity stars, shown along the
    upper horizontal axis, are obtained from the data tables provided
    by \citet{Marigo+01}.  {\it Bottom panel:} The fraction of baryons
    that must be incorporated into stars in order to produce the
    critical flux in the top panel.}
  \label{fig:SingleT}
\end{figure}

How low do the masses of low--mass stars need to be before IR feedback
becomes important?  To answer this question, we next consider stellar
populations with a single effective temperature, in the range $6000
\lsim T_{\rm eff}/{\rm K} \lsim 10^5$, and compute how $J_{\rm crit}$ and
$f_{\rm \ast,crit}$ depend on $T_{\rm eff}$.  The top panel of Figure
\ref{fig:SingleT} shows a dramatic drop in the critical flux owing to
the copious $\sim 2 {\rm eV}$ photon output below $T_{\rm eff} \approx
1.5 \times 10^4$K, or ${\rm M \lsim 2 \msun}$.  As shown in the bottom
panel of this figure, the density in stars required to produce $J_{\rm
crit}$ remains essentially constant down to this $T_{\rm eff}$, and
very near the value required for LW feedback by high--mass stars.
This flatness again results from the cancellation of two effects: the
critical flux and the LW photon output (per stellar baryon) both
decrease significantly as $T_{\rm eff}$ drops below $1.5 \times
10^4$K.

\subsection{Comparison to previous studies}
\label{subsec:others}

To our knowledge, the dominant importance of the radiation from
low--mass stars for global star-formation was not previously discussed
or quantified, with the exception of \citet{Chuzhoy07}.  These authors
evaluated the global impact of ${\rm H^-}$--detachment by an IR
background at later epochs, produced by recombination radiation as the
IGM is becoming significantly ionized (\citealt{Glover07} considered a
similar scenario, with radiation from gas that is highly ionized by
local sources).  They showed that if the ionization is produced by
stars with a soft spectrum, as we consider here, the ${\rm
H^-}$--detachment rate can be further boosted and can become globally
important.  However, they do not explicitly compare ${\rm H^-}$ and LW
feedback, and do not answer the two questions addressed in this {\it
Letter}: how many low--mass stars need to form (1) before ${\rm H^-}$
photo-detachment becomes the dominant feedback mode, and (2) for this
mode to become globally important.

Other previous works have touched on different aspects of the global
radiative feedback discussed here.  Examples include \citet{MBH06} and
\citet{MBH09}, studying radiative feedback in a statistical sample of
several hundred early minihalos in cosmological simulations;
\citet{HAR00} and \citet{JGB08}, studying self-regulation of
star--formation in minihalos through radiative feedback in
semi-analytical models; and \citet{Whalen+10}, studying radiative
feedback in detailed hydrodynamical simulations with the stellar
masses of the sources extending down $25~\msun$.  However, in all of
these works, feedback was due to Lyman-Werner (or ionizing UV) radiation.

\citet{Omukai01}, \citet{BL03}, \citet{SBH10}, \citet{WGH11} and \citet{WHB11} 
have performed calculations similar to the one proposed here, including
${\rm H^-}$ photo-detachment, but have focused on the more massive
($T_{\rm vir} > 10^4{\rm K}$) atomic--cooling halos.  Because the gas in
these halos can cool via neutral H and reach high densities, a much
larger flux $J_{\rm crit}$ is required to suppress ${\rm
H_2}$--cooling. Nevertheless, these works have found results similar
to the ones here: the critical flux for a soft spectrum is $\sim
(1-2)$ orders of magnitude lower than for a hard spectrum.  For
example, using 3D simulations of three different halos, \citet{SBH10}
found $30 < J_{\rm crit}< 300$ ($T_{\rm eff}=10^4$K) versus $10^{4} <
J_{\rm crit}<10^{5}$ ($T_{\rm eff}=10^5$K).  As found here, ${\rm
H^-}$--detachment dominates in the former case, whereas ${\rm
H_2}$--dissociation dominates in the latter.  When self--shielding in
the LW lines of ${\rm H_2}$ is modeled more accurately, however, this
difference is reduced by about an order of magnitude \citep{WHB11}.

\section{Conclusions}
\label{sec:conclusions}

The main conclusion of this {\it Letter} is that if the mass--fraction
of low--mass (few $\msun$) stars exceeded $\sim 90\%$, then their
early IR background radiation dominated over the LW background in
suppressing ${\rm H_2}$ formation. This low-mass fraction is
comparable to those in present-day IMFs, and it is interesting to note
that in this limit, star formation in minihalos could be more
efficient than if the early stars were massive, owing to the lack of
UV feedback and heating inside halos. The early IR background from the
low-mass stars would then exert significant net feedback, and regulate
the star-formation history in the early Universe once a fraction
$f_\ast =$ a few $\times f_{\rm \ast,LW}$ of baryons were converted
into stars. The threshold $f_{\rm \ast,LW}$ is the fraction required
for strong LW feedback (from massive stars), which is $\sim 0.3\%$
that required for reionization assuming ${\rm n_\gamma}= 10$ ionizing
photons per hydrogen atom. Future investigations of radiative feedback
in the early Universe, which include low--mass stars, should therefore
include ${\rm H^-}$ photo-detachment. Our results also highlight the
need for an accurate calculation of the IR photon output of low-mass
stars.

\section*{Acknowledgments}

We thank Greg Bryan and Volker Bromm for insightful discussions, and
Martin Rees, Andrea Ferrara, Daniel Whalen, and Daniel Savin for
useful comments on this manuscript. ZH acknowledges support from NASA
grant NNX11AE05G.


\begin{thebibliography}{59}
\expandafter\ifx\csname natexlab\endcsname\relax\def\natexlab#1{#1}\fi

\bibitem[{{Abel}, {Bryan} \& {Norman}(2002){Abel}, {Bryan}, \&
  {Norman}}]{ABN02}
{Abel} T., {Bryan} G.~L., {Norman} M.~L., 2002, Science, 295, 93

\bibitem[{{Barkana} \& {Loeb}(2001)}]{BLreview}
{Barkana} R., {Loeb} A., 2001, \physrep, 349, 125

\bibitem[{{Bromm}, {Coppi} \& {Larson}(2002){Bromm}, {Coppi}, \&
  {Larson}}]{BCL02}
{Bromm} V., {Coppi} P.~S., {Larson} R.~B., 2002, \apj, 564, 23

\bibitem[{{Bromm}, {Kudritzki} \& {Loeb}(2001){Bromm}, {Kudritzki}, \&
  {Loeb}}]{BKL01}
{Bromm} V., {Kudritzki} R.~P., {Loeb} A., 2001, \apj, 552, 464

\bibitem[{{Bromm} \& {Loeb}(2003)}]{BL03}
{Bromm} V., {Loeb} A., 2003, \apj, 596, 34

\bibitem[{{Caffau} {et~al}\mbox{.}(2011){Caffau}, {Bonifacio}, {Fran{\c c}ois},
  {Sbordone}, {Monaco}, {Spite}, {Spite}, {Ludwig}, {Cayrel}, {Zaggia},
  {Hammer}, {Randich}, {Molaro}, \& {Hill}}]{Caffau+2011}
{Caffau} E. {et~al.}, 2011, \nat, 477, 67

\bibitem[{{Caffau} {et~al}\mbox{.}(2012){Caffau}, {Bonifacio}, {Fran{\c c}ois},
  {Spite}, {Spite}, {Zaggia}, {Ludwig}, {Steffen}, {Mashonkina}, {Monaco},
  {Sbordone}, {Molaro}, {Cayrel}, {Plez}, {Hill}, {Hammer}, \&
  {Randich}}]{Caffau+2012}
---, 2012, A\&A, in press, e-print arXiv:1203.2607

\bibitem[{{Chen} \& {Miralda-Escud{\'e}}(2004)}]{CM04}
{Chen} X., {Miralda-Escud{\'e}} J., 2004, \apj, 602, 1

\bibitem[{{Chuzhoy}, {Kuhlen} \& {Shapiro}(2007){Chuzhoy}, {Kuhlen}, \&
  {Shapiro}}]{Chuzhoy07}
{Chuzhoy} L., {Kuhlen} M., {Shapiro} P.~R., 2007, \apjl, 665, L85

\bibitem[{{Ciardi}, {Ferrara} \& {Abel}(2000){Ciardi}, {Ferrara}, \&
  {Abel}}]{CFA00}
{Ciardi} B., {Ferrara} A., {Abel} T., 2000, \apj, 533, 594

\bibitem[{{Clark} {et~al}\mbox{.}(2011){Clark}, {Glover}, {Klessen}, \&
  {Bromm}}]{Clark+11}
{Clark} P.~C., {Glover} S.~C.~O., {Klessen} R.~S., {Bromm} V., 2011, \apj, 727,
  110

\bibitem[{{Frebel} \& {Norris}(2011)}]{FN11}
{Frebel} A., {Norris} J.~E., 2011, ArXiv e-prints

\bibitem[{{Galli} \& {Palla}(1998)}]{GP98}
{Galli} D., {Palla} F., 1998, \aap, 335, 403

\bibitem[{{Glover}(2007)}]{Glover07}
{Glover} S.~C.~O., 2007, \mnras, 379, 1352

\bibitem[{{Glover} \& {Brand}(2003)}]{GB03}
{Glover} S.~C.~O., {Brand} P.~W.~J.~L., 2003, \mnras, 340, 210

\bibitem[{{Greif} {et~al}\mbox{.}(2011){Greif}, {Springel}, {White}, {Glover},
  {Clark}, {Smith}, {Klessen}, \& {Bromm}}]{Greif+11}
{Greif} T.~H., {Springel} V., {White} S.~D.~M., {Glover} S.~C.~O., {Clark}
  P.~C., {Smith} R.~J., {Klessen} R.~S., {Bromm} V., 2011, \apj, 737, 75

\bibitem[{{Haiman}, {Abel} \& {Rees}(2000){Haiman}, {Abel}, \& {Rees}}]{HAR00}
{Haiman} Z., {Abel} T., {Rees} M.~J., 2000, \apj, 534, 11

\bibitem[{{Haiman}, {Rees} \& {Loeb}(1996){Haiman}, {Rees}, \& {Loeb}}]{HRL96}
{Haiman} Z., {Rees} M.~J., {Loeb} A., 1996, \apj, 467, 522

\bibitem[{{Haiman}, {Rees} \& {Loeb}(1997){Haiman}, {Rees}, \& {Loeb}}]{HRL97}
---, 1997, \apj, 476, 458

\bibitem[{{Haiman}, {Thoul} \& {Loeb}(1996){Haiman}, {Thoul}, \&
  {Loeb}}]{HTL96}
{Haiman} Z., {Thoul} A.~A., {Loeb} A., 1996, \apj, 464, 523

\bibitem[{{Heger} {et~al}\mbox{.}(2003){Heger}, {Fryer}, {Woosley}, {Langer},
  \& {Hartmann}}]{Heger+03}
{Heger} A., {Fryer} C.~L., {Woosley} S.~E., {Langer} N., {Hartmann} D.~H.,
  2003, \apj, 591, 288

\bibitem[{{Hirasawa}, {Aizu} \& {Taketani}(1969){Hirasawa}, {Aizu}, \&
  {Taketani}}]{Hirasawa+69}
{Hirasawa} T., {Aizu} K., {Taketani} M., 1969, Progress of Theoretical Physics,
  41, 835

\bibitem[{{Hirata} \& {Padmanabhan}(2006)}]{HP06}
{Hirata} C.~M., {Padmanabhan} N., 2006, \mnras, 372, 1175

\bibitem[{{Johnson}, {Greif} \& {Bromm}(2008){Johnson}, {Greif}, \&
  {Bromm}}]{JGB08}
{Johnson} J.~L., {Greif} T.~H., {Bromm} V., 2008, \mnras, 388, 26

\bibitem[{{Komatsu} {et~al}\mbox{.}(2011){Komatsu}, {Smith}, {Dunkley},
  {Bennett}, {Gold}, {Hinshaw}, {Jarosik}, {Larson}, {Nolta}, {Page},
  {Spergel}, {Halpern}, {Hill}, {Kogut}, {Limon}, {Meyer}, {Odegard}, {Tucker},
  {Weiland}, {Wollack}, \& {Wright}}]{Komatsu+11}
{Komatsu} E. {et~al.}, 2011, \apjs, 192, 18

\bibitem[{{Kuhlen} \& {Madau}(2005)}]{KM05}
{Kuhlen} M., {Madau} P., 2005, \mnras, 363, 1069

\bibitem[{{Machacek}, {Bryan} \& {Abel}(2001){Machacek}, {Bryan}, \&
  {Abel}}]{MBA01}
{Machacek} M.~E., {Bryan} G.~L., {Abel} T., 2001, \apj, 548, 509

\bibitem[{{Madau} {et~al}\mbox{.}(2004){Madau}, {Rees}, {Volonteri}, {Haardt},
  \& {Oh}}]{Madau+04}
{Madau} P., {Rees} M.~J., {Volonteri} M., {Haardt} F., {Oh} S.~P., 2004, \apj,
  604, 484

\bibitem[{{Marigo} {et~al}\mbox{.}(2001){Marigo}, {Girardi}, {Chiosi}, \&
  {Wood}}]{Marigo+01}
{Marigo} P., {Girardi} L., {Chiosi} C., {Wood} P.~R., 2001, \aap, 371, 152

\bibitem[{{Mesinger}, {Bryan} \& {Haiman}(2006){Mesinger}, {Bryan}, \&
  {Haiman}}]{MBH06}
{Mesinger} A., {Bryan} G.~L., {Haiman} Z., 2006, \apj, 648, 835

\bibitem[{{Mesinger}, {Bryan} \& {Haiman}(2009){Mesinger}, {Bryan}, \&
  {Haiman}}]{MBH09}
---, 2009, \mnras, 399, 1650

\bibitem[{{Mirabel} {et~al}\mbox{.}(2011){Mirabel}, {Dijkstra}, {Laurent},
  {Loeb}, \& {Pritchard}}]{Mirabel+11}
{Mirabel} I.~F., {Dijkstra} M., {Laurent} P., {Loeb} A., {Pritchard} J.~R.,
  2011, \aap, 528, A149

\bibitem[{{Oh}(2001)}]{Oh01}
{Oh} S.~P., 2001, \apj, 553, 499

\bibitem[{{Omukai}(2001)}]{Omukai01}
{Omukai} K., 2001, \apj, 546, 635

\bibitem[{{Omukai} \& {Nishi}(1999)}]{ON99}
{Omukai} K., {Nishi} R., 1999, \apj, 518, 64

\bibitem[{{O'Shea} \& {Norman}(2008)}]{ON08}
{O'Shea} B.~W., {Norman} M.~L., 2008, \apj, 673, 14

\bibitem[{{Prieto} {et~al}\mbox{.}(2011){Prieto}, {Padoan}, {Jimenez}, \&
  {Infante}}]{Prieto+11}
{Prieto} J., {Padoan} P., {Jimenez} R., {Infante} L., 2011, \apjl, 731, L38

\bibitem[{{Ricotti}, {Gnedin} \& {Shull}(2001){Ricotti}, {Gnedin}, \&
  {Shull}}]{Ricotti+01}
{Ricotti} M., {Gnedin} N.~Y., {Shull} J.~M., 2001, \apj, 560, 580

\bibitem[{{Ricotti}, {Gnedin} \& {Shull}(2002){Ricotti}, {Gnedin}, \&
  {Shull}}]{Ricotti+02b}
---, 2002, \apj, 575, 33

\bibitem[{{Ricotti}, {Ostriker} \& {Gnedin}(2005){Ricotti}, {Ostriker}, \&
  {Gnedin}}]{Ricotti+05}
{Ricotti} M., {Ostriker} J.~P., {Gnedin} N.~Y., 2005, \mnras, 357, 207

\bibitem[{{Schaerer}(2002)}]{Schaerer02}
{Schaerer} D., 2002, \aap, 382, 28

\bibitem[{{Schneider} {et~al}\mbox{.}(2012){Schneider}, {Omukai}, {Limongi},
  {Ferrara}, {Salvaterra}, {Chieffi}, \& {Bianchi}}]{Schneider+2012}
{Schneider} R., {Omukai} K., {Limongi} M., {Ferrara} A., {Salvaterra} R.,
  {Chieffi} A., {Bianchi} S., 2012, \mnras, in press, e-print arXiv:1203.4234,
  L444

\bibitem[{{Shang}, {Bryan} \& {Haiman}(2010){Shang}, {Bryan}, \&
  {Haiman}}]{SBH10}
{Shang} C., {Bryan} G.~L., {Haiman} Z., 2010, \mnras, 402, 1249

\bibitem[{{Shapiro} \& {Kang}(1987)}]{SK87}
{Shapiro} P.~R., {Kang} H., 1987, \apj, 318, 32

\bibitem[{{Stacy}, {Greif} \& {Bromm}(2010){Stacy}, {Greif}, \&
  {Bromm}}]{Stacy+10}
{Stacy} A., {Greif} T.~H., {Bromm} V., 2010, \mnras, 403, 45

\bibitem[{{Tumlinson}(2007{\natexlab{a}})}]{Tumlinson07b}
{Tumlinson} J., 2007{\natexlab{a}}, \apj, 665, 1361

\bibitem[{{Tumlinson}(2007{\natexlab{b}})}]{Tumlinson07a}
---, 2007{\natexlab{b}}, \apjl, 664, L63

\bibitem[{{Tumlinson} \& {Shull}(2000)}]{TS00}
{Tumlinson} J., {Shull} J.~M., 2000, \apjl, 528, L65

\bibitem[{{Turk}, {Abel} \& {O'Shea}(2009){Turk}, {Abel}, \&
  {O'Shea}}]{Turk+09}
{Turk} M.~J., {Abel} T., {O'Shea} B., 2009, Science, 325, 601

\bibitem[{{Turk} {et~al}\mbox{.}(2012){Turk}, {Oishi}, {Abel}, \&
  {Bryan}}]{Turk+12}
{Turk} M.~J., {Oishi} J.~S., {Abel} T., {Bryan} G.~L., 2012, \apj, 745, 154

\bibitem[{{Venkatesan}, {Giroux} \& {Shull}(2001){Venkatesan}, {Giroux}, \&
  {Shull}}]{Venkatesan+01}
{Venkatesan} A., {Giroux} M.~L., {Shull} J.~M., 2001, \apj, 563, 1

\bibitem[{{Whalen}, {Hueckstaedt} \& {McConkie}(2010){Whalen}, {Hueckstaedt},
  \& {McConkie}}]{Whalen+10}
{Whalen} D., {Hueckstaedt} R.~M., {McConkie} T.~O., 2010, \apj, 712, 101

\bibitem[{{Whalen} {et~al}\mbox{.}(2008){Whalen}, {O'Shea}, {Smidt}, \&
  {Norman}}]{Whalen+08}
{Whalen} D., {O'Shea} B.~W., {Smidt} J., {Norman} M.~L., 2008, \apj, 679, 925

\bibitem[{{Wise} \& {Abel}(2007)}]{WA07}
{Wise} J.~H., {Abel} T., 2007, \apj, 671, 1559

\bibitem[{{Wise} \& {Abel}(2008{\natexlab{a}})}]{WA08a}
---, 2008{\natexlab{a}}, \apj, 684, 1

\bibitem[{{Wise} \& {Abel}(2008{\natexlab{b}})}]{WA08b}
---, 2008{\natexlab{b}}, \apj, 685, 40

\bibitem[{{Wolcott-Green} \& {Haiman}(2011)}]{WGH11}
{Wolcott-Green} J., {Haiman} Z., 2011, \mnras, 412, 2603

\bibitem[{{Wolcott-Green}, {Haiman} \& {Bryan}(2011){Wolcott-Green}, {Haiman},
  \& {Bryan}}]{WHB11}
{Wolcott-Green} J., {Haiman} Z., {Bryan} G.~L., 2011, \mnras, 418, 838

\bibitem[{{Yoshida} {et~al}\mbox{.}(2003){Yoshida}, {Abel}, {Hernquist}, \&
  {Sugiyama}}]{Yoshida+03}
{Yoshida} N., {Abel} T., {Hernquist} L., {Sugiyama} N., 2003, \apj, 592, 645

\end{thebibliography}

\label{lastpage}

\end{document}